\documentclass[12pt]{iopart}
\usepackage{iopams}
\usepackage{graphics}
\usepackage[next]{inputenc}
\usepackage[dvips]{epsfig}

\begin{document}

\title[Paper submitted to \JPCM]{Buckling instability for a  charged and fluctuating  semiflexible polymer}

\author{Khabat Ghamari and Ali Najafi \footnote[1] {Corresponding author: najafi@znu.ac.ir.}}

\address{Department of Physics, Zanjan University, Zanjan 313, Iran}
\begin{abstract}
In this article we address the problem of Euler's buckling instability in a charged semi-flexible polymer that is under the action of a compressive force. 
We consider this instability as a phase transition and investigate the role of thermal fluctuations in the buckling critical force. 
By performing molecular dynamic simulations, we show that the critical force 
decreases when the temperature increases. Repulsive electrostatic interaction in the finite temperature is in competition with 
thermal fluctuations to increase the buckling threshold.
\end{abstract}



%
%
%

\submitto{\JPCM}


\section{Introduction}
Microtubules and actin filaments are important biopolymers that construct the mechanics of the eukaryotic cells \cite{cell,Nelson,howardbook}. Microtubules the most rigid 
biopolymers of the cellular cytoplasm, organize to 
form asters and mitotic spindle inside  the cells. The dynamics of these structures are essential in most of the physical processes of 
the cell cycle \cite{MTint1,aliee}.  
The actin filaments, the other less rigid biopolymers of the cell with rigidity of about $0.01$ times that of the 
microtubules, are also play various roles in the cell \cite{actinmiosin}. The functioning of these filaments 
in the cell includes, the stability of the cell shape, crawling and the cell movements. Understanding the overall mechanical structure of a 
cell in terms 
of the behavior of the individual constituent polymers is of prime importance in cell mechanics. In addition to the bending properties, 
the buckling instability of the biopolymers in response to the pushing forces is being also observed \cite{MTBuck,MTBuck2}. 
In the case of microtubules two types of the forces initiate the buckling, the forces due to the polymerization when an astral microtubule reaches the 
cell cortex and the forces associated 
with the active molecular motors those are binding the different microtubules to each other or binding an individual microtubule to the cell cortex \cite{howardSpindle,NedelecFluc}. 
Actin filaments also experience the buckling instability because of the interaction with miosin motors \cite{actinmiosin}.
 
The buckling of the biopolymers are subject to either equilibrium or non equilibrium fluctuations. Equilibrium 
fluctuations due to the thermal equilibration with the ambient cytoplasmic fluid and the non equilibrium processes of the 
binding and unbinding of molecular motors to the filaments, enhance the buckling transition. In addition to the fluctuations, 
the other important effect that can change the buckling point, is the electrostatic interactions in a charged  biopolymer. 

These observations show that a complete understanding 
of the mechanics at the cellular scale needs a detail knowledge of the mechanical properties of an individual filament. 
There are some efforts in considering the buckling of either charged or neutral filaments taking into account the fluctuation effects \cite{recent1,recent11,recent2,recent3,kroy,podgornik}.
Here by using the ESPRESSO package \cite{spr}, we perfom the molecular dynamic simulations and focus 
on the numerical simulation for the  buckling instability of either a charged or neutral filament that is subject to equilibrium thermally fluctuations.

The rest of this article is organized as follows: In section 2 we introduce the model with its free energy and physical parameters. In section 3 some known 
analytic mean field results for neutral and charged polymers are summarized. We present the numerical results for a fluctuating filament in the section 4.  

\section{The model}
\begin{figure}
\vskip0.5truecm
\centerline{\epsfxsize=12cm\epsfbox{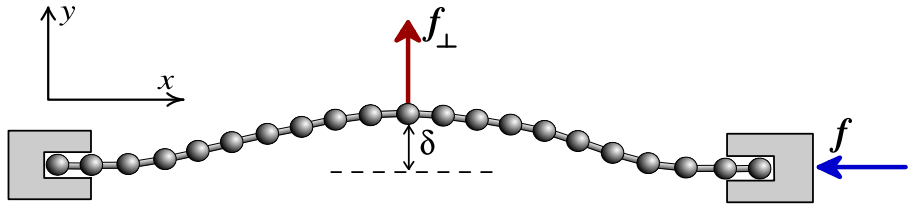}}
\caption{Schematic of a polymer that is under the action of a compression  force $f$, at one of its ends. To investigate the stability of this polymer under this compressional force, we impose a transverse force $f_{\perp}$, at the midpoint of the polymer.}
\label{fig1}  
\end{figure}

Let us consider a stiff polymer with contour length $L$ and bending modulus $\kappa$ embedded in a solution (Fig. \ref{fig1}). 
The space configuration of this polymer is characterized by the 
position vector ${\bf r}(s)$ of a general  points on the chains with contour distance $s$ from one end.
There is a uniform distribution of electric charges on this polymer. 
The free energy of this system has two important counterpart, the bending energy which is due to the chemical bonds 
between the adjacent monomers and the electrostatic repulsion between charges. The free energy of this system reads:
\begin{equation}
G=\frac{\kappa}{2}\int_{0}^{L}ds \frac{\partial^2{\bf r}}{\partial s^2}\cdot \frac{\partial^2{\bf r}}{\partial s^2}+\frac{1}{2}\int_{0}^{L}\int_{0}^{L}
dsds' V(|{\bf r}(s)-{\bf r}(s')|)
+G^{ext},
\end{equation}
where $G^{ext}$ is the contribution to the free energy due to the external forces acting on the chain and  $V({\bf r}(s)-{\bf r}(s'))$, represents the electrostatic repulsion between the charges on the polymer. The electrostatic interaction  
is screened by the mobile counter ions in the solution and is given by \cite{chaikin}:
\begin{equation}
l_{0}^{2}V({\bf r})=\frac{q^2}{\epsilon}~\frac{e^{-r/\lambda_D}}{r},
\end{equation}
where $q$ represents the charge of the ions on the chain that are distributed uniformly with a minimum distance $l_0$, $\epsilon$ stands for the dielectric constant of the solution and  $\lambda_D$ is the Debye-Huckel screening length. The dynamics of this chain in the  over damped, low Reynold's number condition is given by the following equation:
\begin{equation}
\frac{\partial}{\partial t}{\bf r}(s,t)=\left(\frac{1}{\xi_{\parallel}}\hat{t}\hat{t}+\frac{1}{\xi_{\perp}}\hat{n}\hat{n}\right)\cdot
\left(-\frac{\delta{\cal G}}{\delta {\bf r}(s)}+{\bf \xi}(s,t)\right),
\end{equation}
where ${\cal G}$ is the free energy density and the details of the hydrodynamics interactions in a simplified model are collected in two friction coefficients $\xi_{\parallel}$ and $\xi_{\perp}$. 
Here $\xi_{\parallel}$ is the friction coefficient for a small and cylindrical segment of the chain moving along the 
local tangent $\hat{t}$, and 
$\xi_{\perp}$ denotes the corresponding friction for the motion along $\hat{n}$, the unit vector that is  locally normal to the polymer. 
The effects of thermal fluctuations are modeled by a random force ${\bf \xi}(s,t)$ with Gaussian distribution. This means that the 
correlation function for this noisy force reads:
\begin{equation}
\langle {\bf \xi}(s,t){\bf \xi}(s',t') \rangle =2D\delta(s-s')\delta(t-t'),
\end{equation}
where $D$ is the diffusion constant for a small segment of the chain.

The aim of this article is to investigate numerically the response of a chain to a compressive force acting to one end of the chain. Before considering this 
problem we first present the results of some mean field descriptions. 

\section{Mean field results}
We apply a constant compressive force ${\bf f}=-f\hat{x}$ to one end of the polymer. This includes the  following amount of mechanical energy to the free energy:
\begin{equation}
 G^{ext}=-{\bf f}\cdot {\bf r}(s=L),
\end{equation}
minimizing this free energy allows us to study the equilibrium shape of the chain. This can be done for a set of different boundary conditions. Let 
us denote the angle that the local tangent vector at contour length $s$ makes with $x-$axis by $\theta(s)$. 

Two important boundary are the following cases. Case (I): both ends of the chain are clamped ($\theta(0)=\theta(L)=0$) and 
case (II): both ends  can freely 
rotate  ($\dot\theta(0)=\dot\theta(L)=0$). Here the dot symbol represents the derivative with respect to contour length $s$. Physically the 
free rotating end is the case where no external torque is applied to the chain. Considering an electrically neutral chain in the mean field approximation where the effects of thermal fluctuations are not important, one can easily show that the equilibrium shape of the chain is the solution to the following differential equation:
\begin{equation}
\frac{\kappa}{2}\ddot{\theta}(s)+f\sin\theta=0.
\end{equation}
The straight chain solution with $\theta(s)=0$ is a trivial solution that can exist for small forces. But one can see that for forces greater than a 
critical force $f^0_c$, the above solution is not stable and a second solution with buckled shape for the chain initiates. The buckling threshold 
depends on the specific choice of the boundaries. For the  condition where the two ends are clamped, case (I), the bucking force reads: 
$f^0_c=\frac{4\pi^2\kappa}{L^2}$ and for case (II) the buckling instability occurs at $f^0_c=\frac{\pi^2\kappa}{L^2}$.

Taking into account the effects of thermal fluctuation of a neutral semi-flexible chain, enters a new length scale $L_p=\kappa/k_BT$, that is the 
persistence length of a fluctuating semi-flexible polymer. Saddle point analysis for a fluctuating semiflexible polymer under a compressive force is being carried out by T. Odijk \cite{recent2}. It is shown that thermal fluctuations enhances the bucking threshold in the following way:
\begin{equation}
f_c=f^0_c\left(1-c\frac{L}{L_p} \right), 
\label{cnumber}
\end{equation}
where $c$ is a numerical coefficient of the order one. This can be understood as a renormalization of the bending modulus. Thermal fluctuations effectively 
decrease the rigidity of a polymer that is under compressive force. 

To examine the effects of electrostatic interactions in the buckling instability, we note that for a charged polymer some new  
length scales enter into the system. 
For Debye-Huckel electrostatic interaction, two important length scales are the Debye screening length $\lambda_D$ and the 
Bjerrum length $l_B=e^2/\epsilon k_BT$, where $e$ measures the charge of an electron. We assume that ions with charge $Ze$ are distributed uniformly  
on the chain with minimum distance $l_0$. Combining these length scales we can achieve the electric persistence length for this fluctuating charged  polymer 
that is: $L_{pe}=(Z^2/4)(\lambda_D/l_0)^2l_B$. Scaling arguments shows that the total persistence length of a charged chain is given by: 
$L_p+L_{pe}$ \cite{odijk77}. In a very simplified picture for buckling instability in  charged systems we can replace the persistence length in 
the buckling force for a neutral system with the total persistence length: 
\begin{equation}
f_c=f^0_c\left(1-c\frac{L k_BT}{\kappa+\frac{(Ze)^2}{4\epsilon}(\frac{\lambda_D}{l_0})^2} \right)\times\phi(\frac{\lambda_D}{l_0},\frac{\lambda_D}{L}). 
\end{equation}
Based on the dimensional analysis we have considered a general dependence on other length scales through  a dimensionless function $\phi$. This function 
captures the physics of the buckling instability for a charged system at zero temperature. The above expression for the buckling force shows a competition 
between thermal fluctuations and electrostatic repulsion in the buckling instability. Thermal fluctuations decreases the buckling force while the repulsion 
can increase it at finite temperature.

The mean filed theory for the buckling instability in charged polymers are considered in Ref. \cite{recent3}. It is shown that the analysis of the 
buckling problem in the charged polymers with nearly rod like geometry are associated with an harmonic differential equation with a complicated frequency. Defining a new function $u(s)=\dot{\theta}(s)$, and in the limit of $\theta(s)\ll 1$, the mean field equation reads:
\begin{equation}
 \frac{\kappa}{2}\ddot{u}(s)-\omega^2(s)u(s)+fu(s)=0,
\end{equation}
with
\begin{equation}
 \omega^2(s)=\frac{1}{2}\int_{0}^{L}ds'V(|s-s'|).
\end{equation}
The analytical solution of this equation for Debye-Huckel interaction is not possible. Here we consider a short range repulsion interaction with a simple form like: 
\begin{equation}
V(r)=V_0e^{-r/\lambda_D}. 
\end{equation}
In the limit of $L\gg \lambda_D$, we will have $\omega^2\approx2\lambda_DV_0$. Solution to the stability equation, reveals that the 
buckling threshold at zero temperature is: 
\begin{equation}
f_c=f_c^0+\frac{V_0\lambda_D}{l_0^3}.
\end{equation}
This simply shows that the buckling threshold is linearly increasing with increasing the interaction range $\lambda_D$.

\begin{figure}
\vskip0.5truecm
\centerline{\epsfxsize=12cm\epsfbox{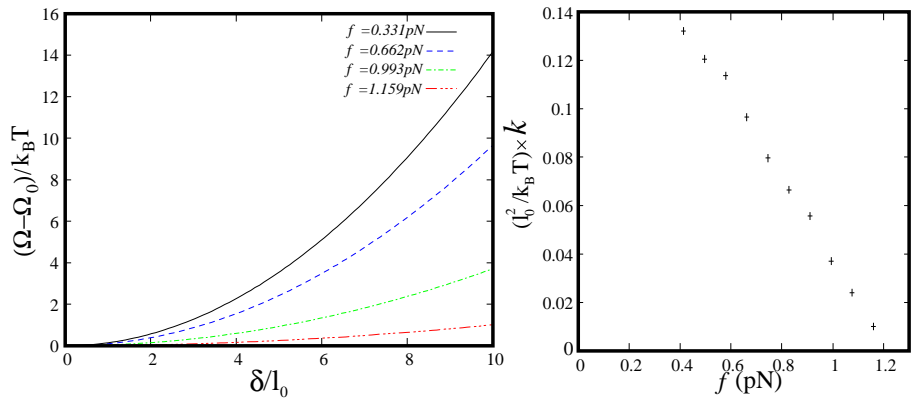}}
\caption{Left: Gibbs free energy $G$, for a buckling polymer versus the perpendicular displacement $\delta$ of 
the polymer at 
its middle point for different values of the external forces $f$. As one can see, for small forces the system is in mechanical stable equilibrium while for large forces the straight state is unstable. Right: the overall stiffness of the midpoint is plotted for different values of the applied 
force. At critical force the stiffness vanishes. Simulations are performed for freely rotating ends.}
\label{fig2}   
\end{figure}

\section{Simulation}
In order to analyze the buckling transition in a charged semiflexible polymer, we 
simulate the system dynamics with ESPResSo software. 
In descretizing the chain, each individual segment has a 
cylindrical geometry with diameter $d=25nm$. For a cylinder with length $l_0$ and diameter $d$, the friction coefficients read \cite{dhont}:
\begin{equation}
 \xi_{\perp}=\frac{4\pi\eta}{\ln (l_0/d)}  ,~~\xi_{\perp}=\frac{2\pi\eta}{\ln (l_0/d)} .
\end{equation}
In our simulations the time step is $dt=1.2 \times 10^{-11}s$, polymer length $L=10\mu m$, number of monomers $N=500$ and the 
viscosity of the solution is $\eta=0.01Pa.s$ (cytoplasmic water). 
In a typical simulation, the equilibration time was about $3\times 10^5 dt$. 

We first perform the simulations for an electrically neutral filament at finite temperature. To investigate the stability of a system that is 
under external compressive force $f$,  we study its response to a small  transverse displacement of a point  in the  middle of the polymer.  
To do this favor we  apply an external auxiliary force ${\bf f}_{\perp}=f_{\perp}\hat{y}$ to the midpoint of the chain. A mechanical work 
with an amount $-f_{\perp}y(s=L/2)$ is included to the free energy. In this case the total free energy of the system is a function of the applied forces:
$G=G(f,f_{\perp})$. Defining a conjugate variable corresponding to the transverse force by, $\delta=-\partial G/\partial f_{\perp}$  we can construct the 
following Legender transformation:
\begin{equation}
\Omega(f,\delta)=G-f_{\perp}\partial G/\partial f_{\perp},
\end{equation}
This is the free energy of the system  that is constrained to a constant $\delta$, the midpoint displacement. Fig. \ref{fig2}(left) shows the numerical 
results of this free energy for different values of the external compressive force $f$. As one can see for small compressive forces the state with 
$\delta=0$ is the stable equilibrium state of the chain. Increasing the compressive force and in a critical force $f_c$, the curvature of the free energy 
changes its sign at point $\delta=0$ signaling the instability transition. To determine the transition point we 
can expand the free energy for small $\delta$ and define  a new variable $k$, that measures 
the  overall stiffness of the midpoint by:
\begin{equation}
\Omega(f,\delta)=\Omega_0(f)+\frac{1}{2}k(f)\delta^2+{\cal O}(\delta)^3,
\end{equation}
the transition point can be reached by  extrapolating the results to find the state with $k(f_c)=0$. 
Fig. \ref{fig2}(right) shows the results for a typical simulation. A numerical value of order $1.2pN$ for the buckling force 
is achieved for a polymer with stiffness $L_p=3mm$. 
This picture for the buckling transition, resembles the physics of the phase transition in magnetic systems with changing the external magnetic 
field. In this case it is very natural to ask what happens if one change the temperature, the strength of the fluctuations. As in case of 
magnetic systems we expect a non trivial change in the transition point. 

\section{Results}
\begin{figure}
\vskip0.5truecm
\centerline{\epsfxsize=14cm\epsfbox{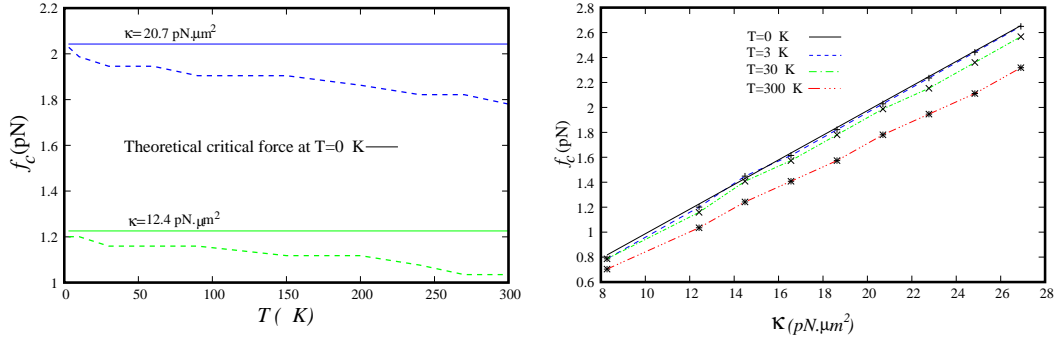}}
\caption{Left: Critical buckling force versus temperature for a neutral semiflexible polymer with two freely rotating ends 
that is allowed to undergo thermal fluctuations. Results are presented for different bending rigidities. Right: 
Buckling force versus bending rigidity is plotted.
In both graphs the 
total length of the polymer is $10~\mu m$, and number of segments in the simulations is $n=500$. Numerical error 
in determining the critical force is $0.04~pN$.}
\label{fig3}   
\end{figure}

To investigate the effects of thermal fluctuations we first consider an electrically neutral filament with bending 
rigidity that has the same order of magnitude of microtubules. For better comparison with cytoplasmic microtubules, we 
choose a total length of $10~\mu m$ and apply the pushing force to one of the ends. For boundary conditions we apply 
the clamped condition to both ends.  
Fig. \ref{fig3}(left) demonstrates the temperature dependence of the critical buckling force for different values 
of the chain stiffness $\kappa$. The results at very low temperature ($T\rightarrow 0$), are in agreement with the mean filed result. 
One can see that by increasing the temperature 
the critical force linearly decreases.   

The temperature dependence of the critical force show different slopes for different numerical values of the 
bending rigidity. To show this behavior in the buckling force we investigated the 
buckling transition at a fixed temperature and vary the rigidity.
Fig. \ref{fig3}(right) shows the behavior of the critical buckling force in terms of the bending stiffness. 
This is investigated for different values of the temperature. 

\begin{figure}
\vskip0.5truecm
\centerline{\epsfxsize=7cm\epsfbox{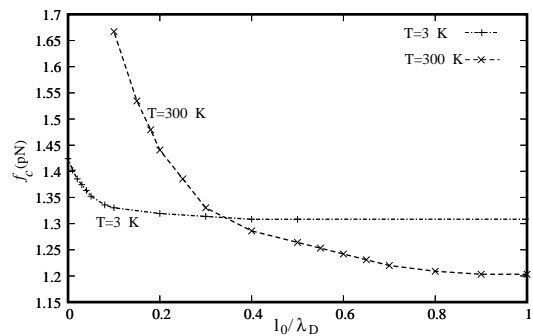}}
\caption{Critical buckling force versus $l_0/\lambda_D$ for a semiflexible and charged polymer with two freely rotating ends at temperatures $3K$ and $300K$ . Here $L_B=0.7nm$, $\kappa=7.45~10^{-26}Nm^2$ (for actin filament) and the total length of the polymer is $0.75\mu~m$.}
\label{fig4}   
\end{figure}

The other issue that we want to address here, is the effects of electrostatic interactions in a charged filament 
immersed in a fluid with mobile charges. For studying the electrostatic effects, we use the Deby-Huckel interaction that 
is already introduced. To have a feeling of a real system, we choose the elastic and electrostatic 
parameters of an actin filament with total length $L=0.75\mu m$ and persistence length $L_p=18\mu m$. Charge density of an actin filament is of 
order $1$ electron in each $2.5A^0$ and the Bjerrum length for water at room temperature is about $0.7nm$. 
Fig \ref{fig4} shows the results of the simulations for buckling critical force in terms of the Debye screening length at two different temperatures.
As one can see, for a range of parameters, there is a competition between the electrostatic effects and the effects due to thermal 
fluctuations. At very small Debye screening length (large $(l_0/\lambda_D)$), the critical force reaches the value that is already 
obtained for a neutral filament. increasing the screening length, the critical force increases. For a constant screening length thermal 
fluctuations can either increase or decrease the critical force.

In summary we have considered the effects due to the thermal fluctuations in the buckling transition for either a charged or 
neutral filaments. We have shown that for a neutral filament thermal fluctuations can always decrease the buckling critical force while for 
a charged filament the effects of fluctuations can compete with the electrostatic effects and either decrease or increase the critical force. 

\section{Acknowledgement}
We would like to acknowledge useful discussion with H. Fazli and M.R. Ejtehadi.

%
%

\end{document}